\documentclass[fleqn,usenatbib,usedcolumn]{mnras}
\usepackage{adjustbox}
\usepackage{newtxtext,newtxmath}
\usepackage[T1]{fontenc}
\usepackage{ae,aecompl}
\usepackage{graphicx}    % Including figure files
\usepackage{amsmath}    % Advanced maths commands

\usepackage{amssymb}    % Extra maths symbols
\usepackage{threeparttable}     %
\usepackage{tikz,xcolor,hyperref}% Make Orcid icon
\definecolor{lime}{HTML}{A6CE39}
\DeclareRobustCommand{\orcidicon}{%
    \begin{tikzpicture}
    \draw[lime, fill=lime] (0,0) 
    circle [radius=0.16] 
    node[white] {{\fontfamily{qag}\selectfont \tiny ID}};   \draw[white, fill=white] (-0.0625,0.095) 
    circle [radius=0.007];  \end{tikzpicture}
    \hspace{-2mm}}
\foreach \x in {A, ..., Z}{%
    \expandafter\xdef\csname orcid\x\endcsname{\noexpand\href{https://orcid.org/\csname orcidauthor\x\endcsname}{\noexpand\orcidicon}}
    }
\def\cha        {{\em Chandra}\/}

\def\twomass    {{2MASS}\/}
\def\vla        {{\em VLA}\/}
\def\sdss       {{\em SDSS}\/}

\def\gmrt       {{\em GMRT}\/}
\def\lofar      {{\em LOFAR}\/}

\title[\cha\ view of Abell~407]{\cha\ view of Abell~407: the central compact group of galaxies and the interaction between the radio AGN and the ICM} 
\author
[Geng et al.]{Chao Geng$^{1,2}$\thanks{gengchao@pmo.ac.cn}\orcidE{}, Chong Ge$^{1}$\thanks{astrochongge@gmail.com}\orcidC{}, Dharam V. Lal$^{3}$\orcidD{}, Ming Sun$^{4}$\thanks{ming.sun@uah.edu}\orcidA{}, Li Ji$^{1}$\orcidB, 
\newauthor
Haiguang Xu$^{5}$,
Wenhao Liu$^{1}$\orcidJ{},
Martin Hardcastle$^{6}$\orcidF{}, William Forman$^{7}$\orcidI{},
\newauthor
Ralph Kraft$^{7}$\orcidH{},
Christine Jones$^{6}$\orcidG{}\\
$^{1}$Purple Mountain Observatory, Chinese Academy of Science, Nanjing  210034, People's Republic of China\\
$^{2}$Department of Astronomy, University of Science and Technology of China, Hefei 230026, People's Republic of China\\
$^{3}$National Centre for Radio Astrophysics, Tata Institute of Fundamental Research,  Post Box 3, Ganeshkhind P.O., Pune 411007, India\\
$^{4}$Department of Physics and Astronomy, University of Alabama in Huntsville, Huntsville, AL 35899, USA\\
$^{5}$School of Physics and Astronomy, Shanghai Jiao Tong University, 800 Dongchuan Road, Minhang, Shanghai 200240, People's Republic of China\\
$^{6}$Centre for Astrophysics Research, School of Physics, Astronomy $\&$ Mathematics, University of Hertfordshire, College Lane, Hatfield AL10 9AB, UK\\
$^{7}$Harvard-Smithsonian Center for Astrophysics, 60 Garden Street, Cambridge, MA 02138, USA\\
}

\begin{document}
\date{Accepted. Received; in original form}

\pubyear{2021}

\maketitle

\begin{abstract}
        % basic of A407. 
        Abell 407 (A407) is a unique galaxy cluster hosting a central compact group of nine galaxies (named as ‘Zwicky’s Nonet’; G1 - G9 in this work) within a 30 kpc radius region. 
        The cluster core also hosts a luminous radio active galactic nucleus (AGN), 4C~35.06 with helically twisted jets extending over 200 kpc.
        With a 44 ks \cha\ observation of A407, we characterize the X-ray properties of its intracluster medium (ICM) and central galaxies. The mean X-ray temperature of A407 is 2.7 keV and the $M_{200}$ is $1.9 \times 10^{14}\, {M_{\odot}}$. 
        We suggest that A407 has a weak cool core at $r < 60$ kpc scales and at its very center, $< 1$-2 kpc radius, a small galaxy corona associated with the strong radio AGN.
        We also conclude that the AGN 4C~35.06 host galaxy is most likely G3.
        We suggest that the central group of galaxies is undergoing a `slow merge' procedure. The range of the merging time-scale is $0.3\sim2.3$ Gyr and the stellar mass of the future brightest cluster galaxy (BCG) will be $7.4\times10^{11} M_{\odot}$. 
        We find that the regions which overlap with the radio jets have higher temperature and metallicity. This is consistent with AGN feedback activity.
        The central entropy is higher than that for other clusters, which may be due to the AGN feedback and/or merging activity.
        With all these facts, we suggest that A407 is a unique and rare system in the local universe that could help us to understand the formation of a massive BCG.
\end{abstract}

\begin{keywords}
galaxies: clusters: individual: Abell 407 -- galaxies: clusters: intracluster medium -- X-rays: galaxies: clusters -- galaxies: ISM -- galaxies: elliptical and lenticular, cD
\end{keywords}

\section{Introduction}
Galaxy clusters are the most massive virialized systems in the Universe. They form hierarchically through the merger of smaller substructures \citep{2005RvMP...77..207V}. 
Clusters not only provide cosmological constraints but also provide a test of structure formation scenarios \citep{2012ARA&A..50..353K}. 
The cores of massive clusters are one of the densest galaxy environments in the Universe, and thus are expected to host the strongest dynamical evolution processes.
The brightest cluster galaxy (BCG) is usually located at the minimum potential well of the cluster and lies close to the peak of the X-ray emission \citep{1983ApJ...274..491B,1984ApJ...276...38J}. They are the largest and most massive galaxies in the Universe and usually an order of magnitude brighter than typical elliptical galaxies \citep{2007MNRAS.379..867V}. They may also host the most massive super-massive black holes (SMBHs) in the Universe \citep{2011Natur.480..215M,2016Natur.532..340T}.
Hence, they have a distinctive formation history \citep{1978ApJ...223..765D}. 

There are several different formation scenarios for BCGs: 
(1) the continuous merging of galaxies in the cluster potential, known as `galactic cannibalism'  \citep{1975ApJ...202L.113O,1978ApJ...224..320H}; 
(2) direct cooling from the hot-gas halo onto a preexisting central galaxy \citep{1977ApJ...215..723C,1994ARA&A..32..277F};
(3) tidal stripping of cluster galaxies that pass near the cluster center \citep{2012MNRAS.427.2047T};
(4) early merging during cluster collapse \citep{1985ApJ...289...18M,1990dig..book..394T,1994MNRAS.268...79W,1998ApJ...502..141D}. 
However, previous works have shown that mergers of smaller non-star-forming early-type galaxies (so-called `dry mergers') are the primary mechanism through which BCGs grow from $z \sim 1$ to the present day \citep{2005AJ....130.2647V,2006ApJ...640..241B,2007MNRAS.375....2D}.
\cite{2009MNRAS.396.2003L} find that the merging rate increases with both the galaxy mass and age, which leads to the large fraction of dry mergers in the nearby universe. \cite{2009ApJ...700L..21K} suggest there are about $10\%\sim20\%$ of massive galaxies with a dry merger in the last Gyr.

\begin{figure*} 
        \centering 
        \includegraphics[width=\textwidth,keepaspectratio=true,clip=true]{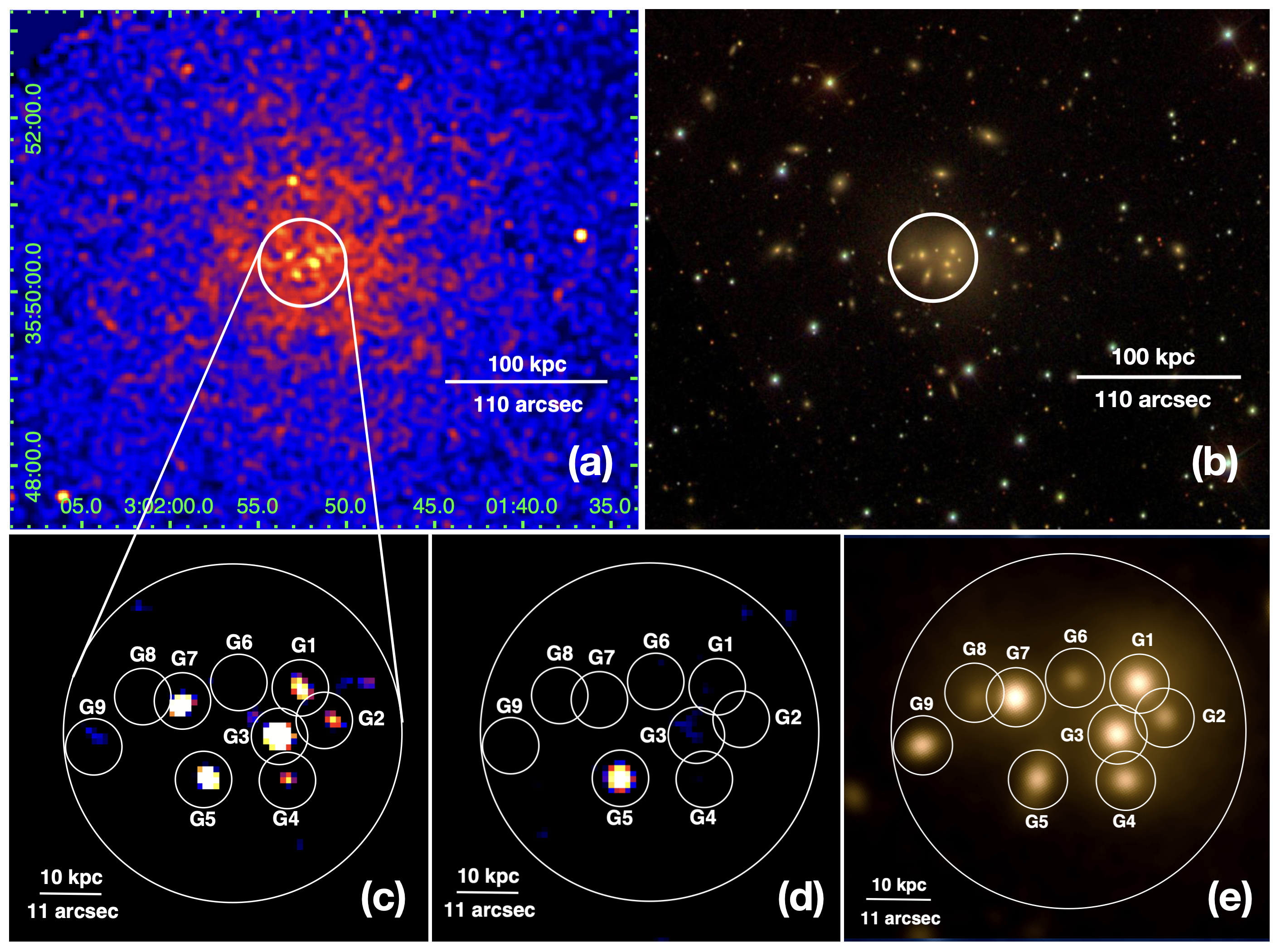} 
        \caption{ 
        X-ray and optical images of Abell 407 and its central region. The central region hosts nine closely packed galaxies which are embedded within a diffuse stellar halo. This system is possibly a rare cD galaxy precursor assembling in a galaxy cluster environment. 
            {\it (a)}: \cha\ background-subtracted and exposure corrected image of A407 in the 0.5 - 2 keV band. 
            {\it (b)}: \sdss\ three-color composite image (blue: $g$-band; green: $r$-band; red: $i$-band) of the same field as the upper left panel. 
            {\it (c)}: Soft X-ray (0.5-2 keV) band image of the central galaxies within a circle of 30$''$ radius. 
            {\it (d)}: Hard X-ray (2-7 keV) band image of the central galaxies. 
            {\it (e)}: \sdss\ three-color composite image of the central galaxies. 
        All the X-ray images are binned with a size of two 0.492$''$ pixels and smoothed with a uniform Gaussian kernel of 1.97$’'$.
            } 
        \label{pic overview}
    \end{figure*}

Abell 407 (A407; $z = 0.0462$) contains a compact group of elliptical galaxies in a luminous matrix (Fig.~\ref{pic overview}). 
Several bright galaxies appear to be embedded in a diffuse optical halo within a region of diameter $\sim$60 kpc \citep{1982ApJ...263...14S}. 
It possibly provides a unique laboratory for the ongoing formation of a giant cD galaxy. 
In addition, the BCG-hosted active galactic nucleus (AGN) has a strong influence via `AGN feedback' on its surroundings. In the radio bands, A407 shows 400-kpc scale helically twisted jets \citep{2015A&A...579A..27S,2017MNRAS.471..617B}. Thus, A407 also provides a great opportunity to study the interaction between the galaxy nuclear activity and the cluster environment. 

In this article, we make a detailed study of this remarkable system. 
This paper is organized as follows. In Section~\ref{sec:previous_observations}, we summarise the previous optical and radio observations of A407. Section~\ref{sec:data reduction} presents the \cha\ data reduction methods. Section~\ref{sec:results} is the X-ray observation results. Section~\ref{sec:discussion} is the discussion of the results.
Section~\ref{sec:summary} is our summary.
Throughout this article, we assume a cosmology with $H_0$ = 70 km s$^{-1}$ Mpc$^{-1}$, $\Omega_{\rm m}=0.3$, and $\Omega_{\Lambda}= 0.7$. The redshift of A407 $z=0.0462$ gives $1^{\prime\prime}=0.907$ kpc. Uncertainties quoted in this paper are 1$\sigma$.

\begin{table*}
        \centering
        \caption{Galaxy properties in the central region of Abell 407 }
        \tabcolsep=0.1cm
        \begin{tabular}{@{}lcccccccccc@{}}
            \hline\hline
            Galaxy$^a$                  & RA, DEC            & cz$^b$           & $L_{Ks}$$^c$ & Mass$^d$ & SMBH Mass$^e$ & Net counts    			 & log$L$     				 	 & log$L$      &  HR$^f$                  & log$L_{X}$$^g$   \\ 
                                        & (J2000)            & (${\rm km\ s^{-1}}$)     & ($10^{10}$$L_{\odot}$) &   ($10^{10}$$M_{\odot}$) & ($10^{8}$$M_{\odot}$)  & (0.5-7keV)    &    (0.5-2keV)            &       (2-7keV)         &       			 &  (0.5-2keV)       \\ 
            \hline
            G1        & 03h 01m 51.5s +35d 50m 28.2s & 14339 (481)  &8.5          & 12.63       &  $ 3.88 \pm 1.23 $       & $23  \pm 8$             & $40.02_{-0.33}^{+0.19}$       & $39.60_{-0.33}^{+0.19}$      & $-0.61_{-0.31}^{+0.31}$  & 40.82  \\ 
            G2        & 03h 01m 51.2s +35d 50m 22.0s & 13600 (258)  &2.7          & 3.07        &  $ 0.53 \pm 0.44 $       & $14  \pm 7$             & $40.25_{-0.24}^{+0.15}$       & $39.83_{-0.24}^{+0.15}$      & $<-0.16$                 & 39.07  \\ 
            G3        & 03h 01m 51.8s +35d 50m 19.6s & 14119 (261)  &8.1          & 10.71       &  $ 9.83 \pm 2.96 $       & $41  \pm 9$             & $40.41_{-0.13}^{+0.10}$       & $39.99_{-0.13}^{+0.10}$      & $-0.46_{-0.20}^{+0.20}$  & 40.74  \\ 
            G4        & 03h 01m 51.7s +35d 50m 12.0s & 15030 (1172)  &4.7          & 6.32        &  $ 0.40 \pm 0.45 $       & $13  \pm 7$             & $39.29_{-0.18}^{+0.13}$       & $38.87_{-0.18}^{+0.13}$      & $-0.39_{-0.52}^{+0.52}$  & 39.93  \\ 
            G5        & 03h 01m 52.9s +35d 50m 12.0s & 13199 (658)  &2.2          & 2.95        &  $ 4.52 \pm 0.74 $       & $42  \pm 9$             & $40.38_{-0.12}^{+0.09}$       & $40.51_{-0.11}^{+0.09}$      & $ 0.21_{-0.21}^{+0.21}$  & 38.79  \\ 
            G6        & 03h 01m 52.4s +35d 50m 28.9s & 13299 (558)  &0.6          & 0.57        &  $ 0.52 \pm 0.65 $       & $<19      $             & $<40.68$                      & $<41.06$                     & --                       & 36.73  \\ 
            G7        & 03h 01m 53.2s +35d 50m 25.7s & 13990 (132)  &7.6          & 9.96        &  $ 3.08 \pm 1.00 $       & $17  \pm 7$             & $40.07_{-0.28}^{+0.17}$       & $39.65_{-0.28}^{+0.17}$      & $<-0.25$                 & 40.65  \\ 
            G8        & 03h 01m 53.8s +35d 50m 26.5s &  --          & --          & --          &  --       & $<17      $             & $<40.72$                      & $<40.84$                     & --                       & --     \\ 
            G9        & 03h 01m 54.5s +35d 50m 17.7s & 13520 (338)  &1.9          & 2.22     & $ 1.34 \pm 0.68 $   &   $12 \pm 7$       & $39.99_{-1.61}^{+0.30}$      & $39.43_{-1.61}^{+0.30}$      & $<-0.05$                 & 38.56  \\ 
            \hline            
        \end{tabular}
        \begin{tablenotes}
        \item $^a$  Galaxy names are the same as \cite{2017MNRAS.471..617B}. %Notations in the parentheses are resolved from NED.
        \item $^b$  Galaxy velocity values are from NED. The values in parentheses are the absolute value of difference between the galaxy velocities and the HI absorption velocity (13858${\rm km\ s^{-1}}$) detected by \cite{2015A&A...579A..27S}. The HI gas has a similar velocity to that of the host galaxy for 4C 35.06.
        \item $^c$  The $K_s$ band luminosity is derived from the \twomass\ point source catalog (PSC). We also measure the total $K_s$ band luminosity within 35$''$ radius of the cluster center, which contains all nine galaxies and their extended emission. The total luminosity is $8.52\times10^{11} L_{\odot}$.  %$\log \left(L_{K} / L_{\odot}\right)$ (extinction corrected).  The unit is $10^{10}$ $L_{\odot}$. 
        \item $^d$  The galaxy stellar masses are derived from $L_{Ks}$ with the stellar mass-to-light ratio $\log _{10}(M / L)=a_{\lambda}+b_{\lambda} \times \text { color }$ \citep{2003ApJS..149..289B}. 
        \item $^e$ The SMBH masses are calculated by \cite{2017MNRAS.471..617B} from bulge stellar velocity dispersion. 
        \item $^f$  HR = (H - S)/(H + S), HR: Hardness ratio; H: count rate at the 2 - 7 keV band; S: count rate at the 0.5 - 2 keV band. 
        \item $^g$  The predicted log($L_{\rm 0.5 - 2 keV}$ / erg s$^{-1}$) from the $L_{Ks}$-$L_{X}$ relation \citep{2007ApJ...657..197S}.
        \end{tablenotes}
        \label{table galaxies}
\end{table*}

\section{Previous Observations} % (fold)
\label{sec:previous_observations}

\subsection{Optical detection} % (fold)
\label{sub:optical_detection} 
Fig.~\ref{pic overview}(b) shows the optical image of the central region of A407 from the Sloan Digital Sky Survey (\sdss). It shows a complex ensemble of at least nine galaxies within $\sim$1 arcmin. The central galaxies are embedded in a diffuse and low surface brightness stellar halo. Table~\ref{table galaxies} lists the optical $K_s$ band magnitudes, line of sight velocities, and masses of these nine galaxies.

The compact group of galaxies in A407 was first noticed by \cite{1971cscg.book.....Z}. \cite{2017MNRAS.471..617B} proposed to name this extraordinary system as `Zwicky's Nonet'.
It was later studied in more detail by \cite{1982ApJ...263...14S} who suggest this puzzling system is a snapshot of a forming cD galaxy.
They consider that it is unlikely to be a pre-existing central cD capturing multiple galaxies. Instead, the cD is being created as a result of tidal disruption/merger. In summary, a recent merger generated the large-scale envelope and dynamical friction caused the remaining galaxies to merge.

\subsection{Radio detection} % (fold)
\label{sub:radio_detection}
The compact group also hosts a moderately luminous radio source, 4C~35.06. The earliest detection of this source was at 1.4 GHz with the Cambridge one-mile telescope \citep{1975MNRAS.170...53R}. \cite{1993A&AS..101..431B} studied this source with the Very Large Array (\vla) at 1.4 $\&$ 5 GHz and found a structure with two lobes. The Very Long Baseline Array ({\em VLBA}) at 5 GHz found that the large-scale radio galaxy is most likely to be identified as the inner galaxy G3 \citep{2010A&A...516A...1L}. 
Later, \cite{2015A&A...579A..27S} studied this source with the Low-Frequency Array (\lofar) at 62 MHz and found the radio jets have a helical morphology. \cite{2017MNRAS.471..617B} has also taken multifrequency radio observations of this source with the Giant Meter-wave Radio Telescope (\gmrt) at 610, 235, and 150 MHz. They revealed that the large-scale radio structures show 400-kpc scale helically twisted radio jets and outer diffuse lobes. 

The radio observations present an interesting helically twisted radio jet structure that could be triggered by the merger of galaxies. To reveal more details of the merging progress and identify the progenitor galaxy of the large-scale jet structure firmly, higher spatial resolution radio or X-ray data are still needed. X-ray observations can also help us to study the interaction signatures between the hot gas and the radio AGN.

\section{Observation and Data Reduction} % (fold)
\label{sec:data reduction}
The X-ray observation used in this paper was performed with the \cha\ Advanced CCD Imaging Spectrometer (ACIS). The observation ID is 18267 (PI: Dharam V. Lal). The exposure time is 43.5 ks. We use \cha\ Interactive Analysis of Observation (CIAO, version 4.11) and calibration database (CALDB, version 4.7.7) to reprocess the data following a procedure similar to that described in \cite{2009ApJ...693.1142S} and \cite{2019MNRAS.484.1946G}. 
We reprocess a new level 2 event file using the {\sc chandra\_repro} script with the VFAINT mode correction. 
We investigate the light curve in the band $0.5\sim9.5$ keV from source-free regions to remove the intervals affected by flares, using the {\sc deflare} script. Such intervals are defined as those deviating more than 3$\sigma$ from the mean rate. We use the {\sc wavdetect} script to detect discrete sources with the point spread function (PSF) map created by the {\sc mkpsfmap} script. Blank-sky background files are processed with the {\sc blanksky} script. 

To obtain the background-subtracted and exposure corrected cluster X-ray image, we scale the blank-sky background for each CCD according to the exposure time and count rate in the 9.5-12 keV band. The corresponding exposure maps are generated with the {\sc fluximage} script. The final X-ray image is shown in Fig.~\ref{pic overview}.

As explained in the Appendix of \cite{2009ApJ...693.1142S}, there are two basic components in the quiescent \cha\ background, the instrumental background, and the cosmic X-ray background (CXB), which are not separately recorded in the blank-sky background file. 
The CXB would be unphysically scaled if we scale the blank-sky spectrum according to the flux of the instrumental background. 
After subtracting the scaled blank-sky background, there would still be some residual X-ray sky background in the spectrum. 

To get a clean spectrum of the cluster, we use the method of `double subtraction' to remove the residual sky background. 
Based on the surface brightness profile (SBP) in Fig.~\ref{pic sbp}, we first choose an off-source region in ACIS-I3 where the cluster emission is negligible and the sky background dominates. 
Then the spectra are extracted with {\sc specextract} script for both the off-source region and the cluster region. The blank-sky background is rescaled according to the count rate in the 9.5-12 keV band. 
Finally, the spectra of the off-source region and the cluster region are jointly fitted with the {\tt TBABS*(APEC+PL)+APEC} model, with the normalizations connected by the ratio of their areas. The {\tt APEC} model with absorption represents the cluster emission component. The power-law model with absorption represents the cosmic hard X-ray background composed of unresolved X-ray point sources \citep{2009ApJ...693.1142S}. The second {\tt APEC} model without absorption represents the residual sky background \citep{2005ApJ...628..655V}. 
The spectra are fitted jointly with the XSPEC package. We use the AtomDB (version 3.0.8) database of atomic data and the solar abundance tables from \cite{1989GeCoA..53..197A}.  
For the hydrogen column density $N_{\rm H}$, we adopt the value $1.72 \times 10^{21} \rm{cm^{-2}}$ from the NHtot tool \citep{2013MNRAS.431..394W}. 

\section{Analysis and Results} % (fold)
\label{sec:results}
   
\subsection{X-ray properties of central galaxies}
\label{subsec:galaxies}
%X-ray image
Fig.~\ref{pic overview}(a) is the background-subtracted and exposure corrected image of A407 in the 0.5-2.0 keV band. The central galaxies in a region with a radius of $30''$ are shown in Fig.~\ref{pic overview}(c,d). We use the circles with a $5''$ radius to mark the location of each galaxy. Fig.~\ref{pic overview}(b) is the corresponding \sdss{} image. 
There are seven out of nine optical bright galaxies detected in the soft X-ray (0.5-2.0 keV) image. Among the seven galaxies, the soft X-ray emission from G1, G2, G4, G7, and G9 is weak, while G3 and G5 host the brightest soft X-ray sources in the group. In the hard X-ray (2.0-7.0 keV) image in Fig.~\ref{pic overview}(d), G5 is the only X-ray bright galaxy.

We derive net X-ray counts for the galaxies in both soft and hard X-ray bands. 
For the count measurement, we use the local background to exclude the contribution from the surrounding hot intracluster medium (ICM). 
Uncertainties for net counts are given as $1 \sigma$ and upper limits are given as $3 \sigma$.
The results are listed in Table~\ref{table galaxies}. 
As shown in Table~\ref{table galaxies}, only G6 and G8 are not detected.
We give the upper limits of their net counts and exclude them in the following HR and spectral analysis.

From the counts of the central galaxies, we calculate the hardness ratio for each galaxy. The hardness ratio is defined as HR = (H - S)/(H + S), where S and H are the net counts in the soft and hard X-ray bands \citep{2004ApJ...612L.109W}. The results are shown in Table~\ref{table galaxies}.
In the hard X-ray band, since the counts of G2, G7 and G9 are close to the background, we could only give the upper limits of their HR.
G5 is the only source that has more counts in the hard X-ray band than the soft band, which may suggest intrinsic absorption for its AGN. 
The X-ray emission of other galaxies is soft. 

We combine the spectra of G1, G2, G3, G4, G7, and G9 as they have similar hardness ratios. 
We fit the combined spectra with an absorbed {\tt APEC} thermal model.
The redshift and the abundance we adopt are 0.0462 and 1 solar value \citep{1982ApJ...263...14S}. 
The fitting result shows that the mean temperature of the X-ray emission from these six galaxies is $1.6^{+1.2}_{-1.3}$ keV.
To examine the potential contribution from AGNs and LMXBs,  we also add a power-law component with a photon index of 1.7 to fit the combined spectra, as did in \cite{2007ApJ...657..197S}. We find the 0.5-2 keV luminosity of the power-law component is at most 30\% of the total 0.5-2 keV luminosity,
which suggests that most of the flux arises from thermal coronae.  

Due to the limited counts of each galaxy, we estimate their luminosities by fitting a model with fixed parameters. The luminosities are given both in the soft and hard X-ray bands. For G1, G2, G3, G4, G7, and G9, we use the single absorbed {\tt APEC} model with a temperature of 1.6 keV and 1 solar abundance. For G5, we adopt the single absorbed power-law model with the typical AGN photon index of 1.7. For G6 and G8, we give the estimated upper limits. All results are listed in Table~\ref{table galaxies}. 

\subsection{X-ray properties of cluster hot gas} 
\label{sub:4.2hotgas} 

\begin{figure}
        \centering
        \includegraphics[angle=0,width=0.47\textwidth]{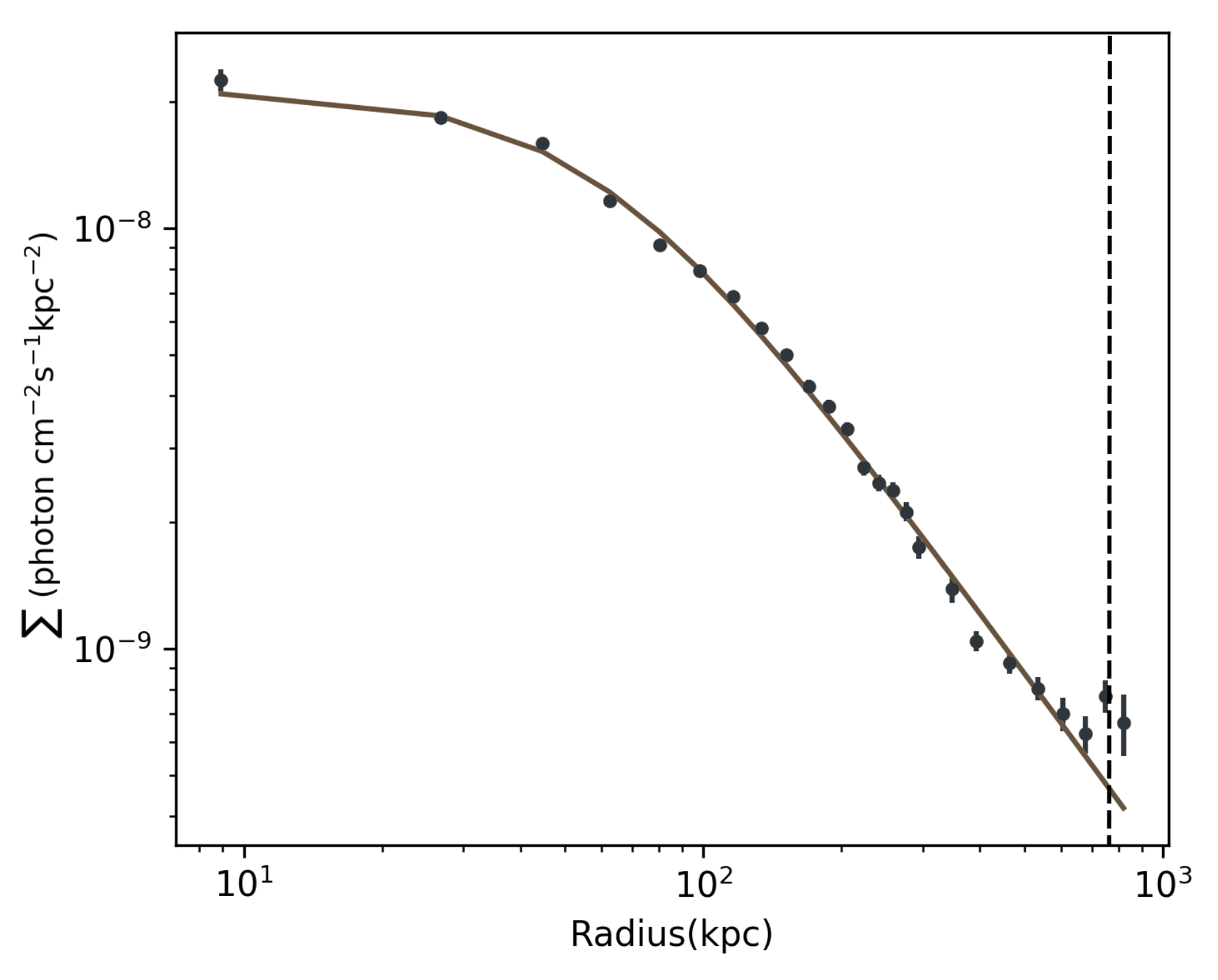}
        \caption{Surface brightness profile of A407 in 0.5-2 keV band. The dashed line marks the $r_{500}$ of A407, which is 774 kpc. The brightness profile is fitted with the $\beta$-model. 
        The best fitting parameters are listed in Table~\ref{tab:beta_label}.
        }
        \label{pic sbp}
    \end{figure}

\begin{figure*}
        \centering
        \includegraphics[width=0.48\textwidth,keepaspectratio=true,clip=true]{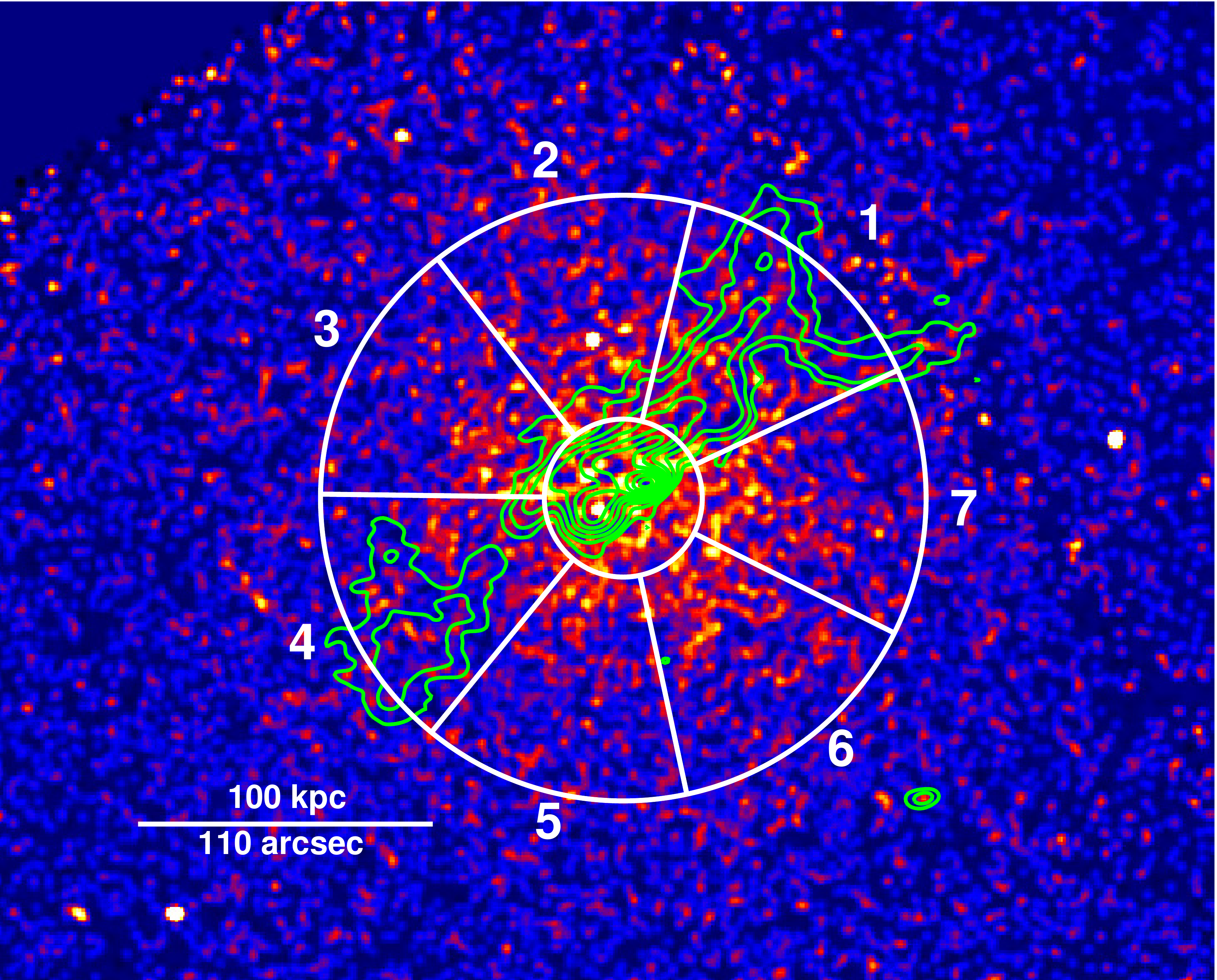}
        \includegraphics[width=0.48\textwidth,keepaspectratio=true,clip=true]{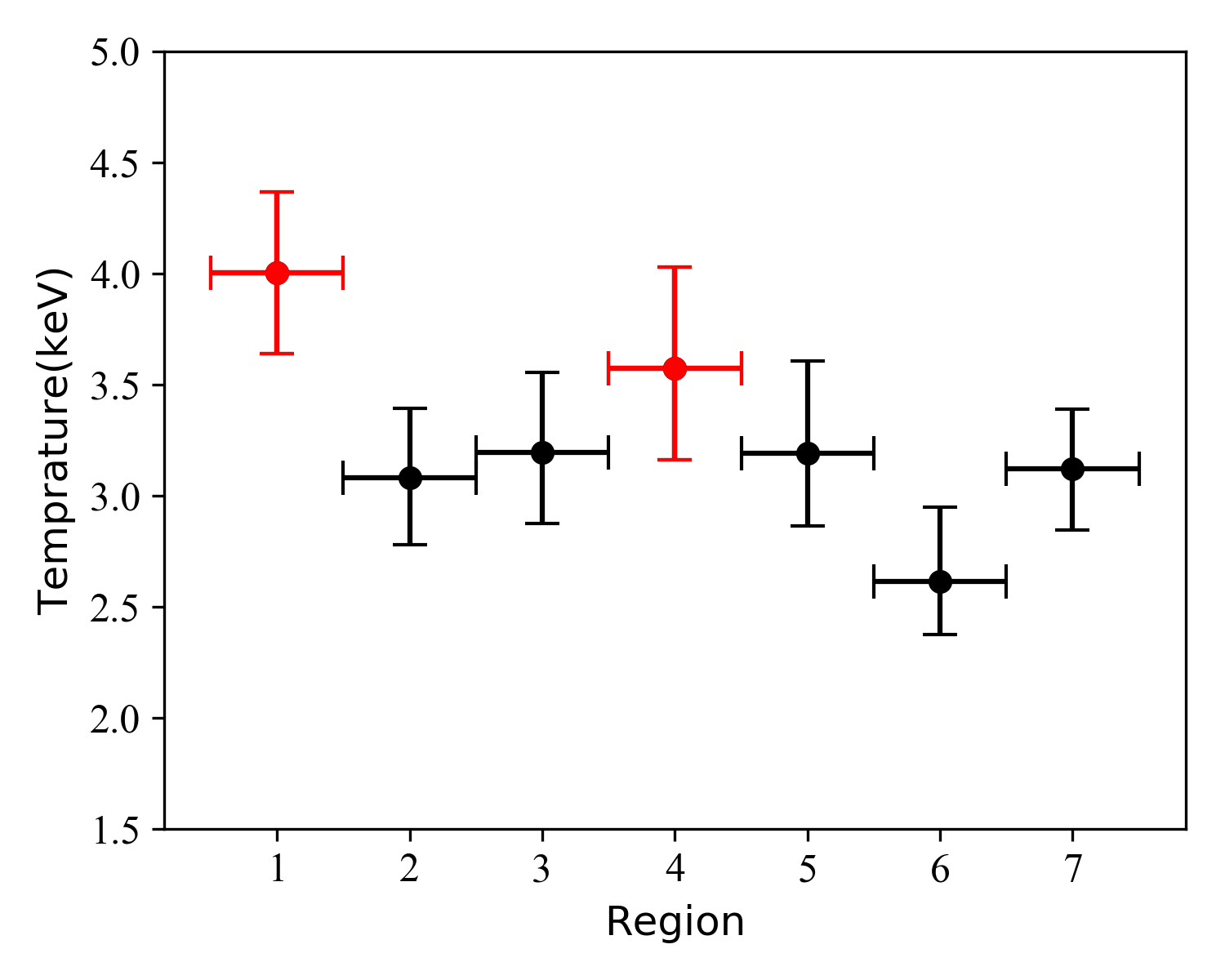}
        \caption{Fitting results of seven panda regions. Left-hand panel: \cha\ 0.5-2 keV image overlaid with \gmrt\ 610 MHz contours. 
        The X-ray image is binned with a size of two 0.492$''$ pixels and smoothed with a uniform Gaussian kernel of 1.97$’'$.
        The inner radius of the sector regions is 30$''$ and the outer radius is 115$''$. Right-hand panel: the temperature is plotted against azimuthal region number. The regions that contain the radio jets (plotted in red) have the maximum temperatures which are $4.00_{-0.36}^{+0.36}$ and $3.57_{-0.41}^{+0.45}$ keV respectively. The temperature and abundance values are listed in Table~\ref{tab:panda_label}. 
        } 
        \label{pic panda result} 
    \end{figure*}

We extract a SBP from the ICM in Fig.~\ref{pic sbp}. SHERPA is used to fit the SBP with the $\beta$-model \citep{1976A&A....49..137C}: 
%beta model: 
\begin{equation} 
    I=I_{0}\left(1+x^{2}\right)^{1 / 2-3 \beta}, 
\end{equation}
where $x=R / r_{\mathrm{c}}$, and $I_{0}, r_{c}$, and $\beta$ are free parameters. The ICM distribution can be reasonably fit by the $\beta$-model, which is shown in Fig.~\ref{pic sbp}. The best-fitting parameters are in Table~\ref{tab:beta_label}. We also check the SBP in the sector regions shown in Fig.~\ref{pic panda result} and the SBPs are similar in different directions. 

%sbp profile 

\begin{table}
    \centering
    \caption{The $\beta$-model fitting results of A407's surface brightness profile}
    \begin{tabular}{l|c}
    \hline\hline 
        Parameters & Value \\
        \hline
        $I_{0}$ ($10^{-8} \text { photons } \mathrm{s}^{-1} \mathrm{cm}^{-2} \operatorname{arcsec}^{-2}$) & $1.75 \pm 0.07$  \\
        $r_{c}$ (arcsec) & $66.7 \pm 2.9$  \\
        $\beta$ & $0.42 \pm 0.01$ \\
    \hline
    \end{tabular}
    \label{tab:beta_label}
\end{table}    

\begin{figure*}
        \centering
        \includegraphics[width=0.48\textwidth,keepaspectratio=true,clip=true]{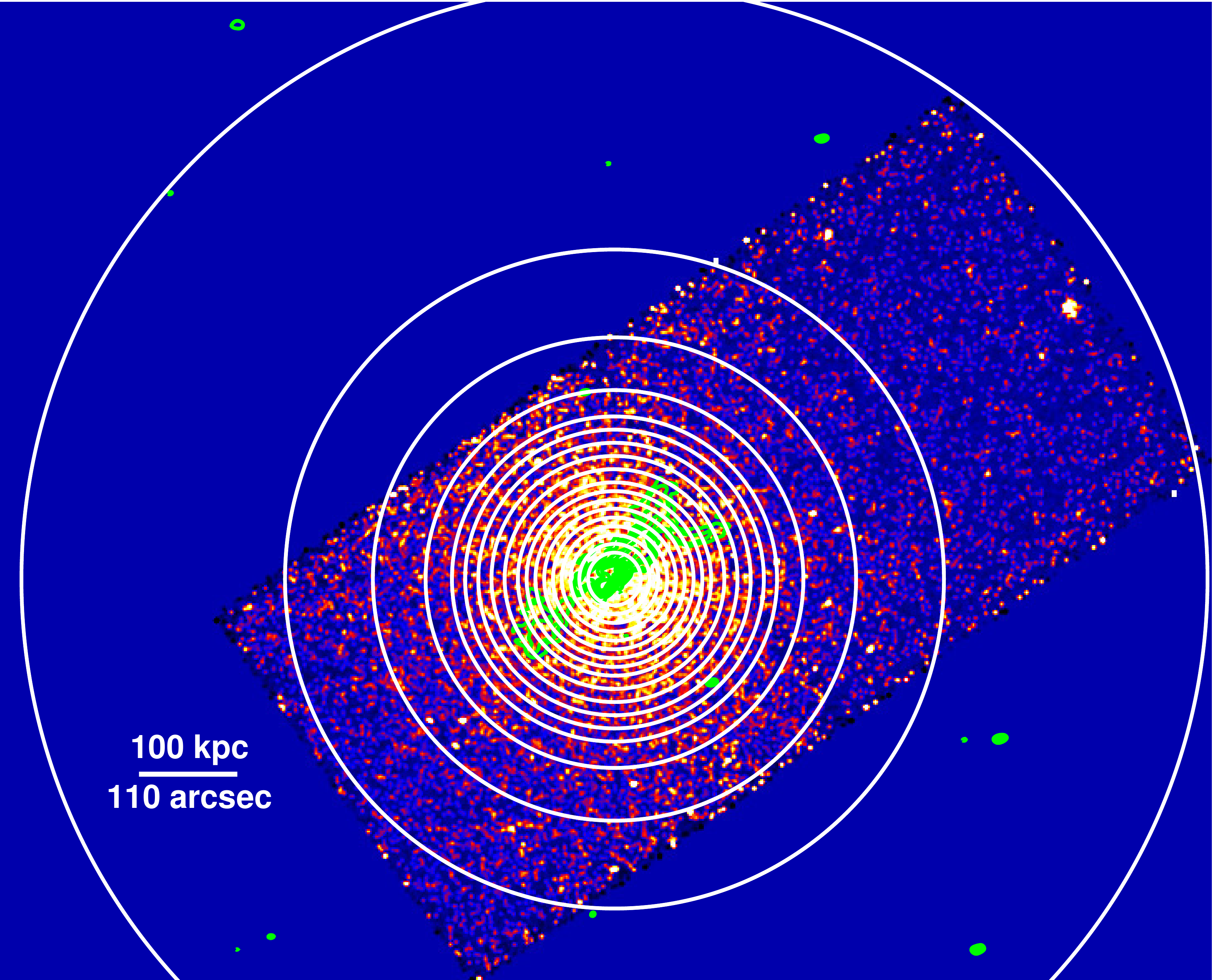}
        \includegraphics[width=0.48\textwidth,keepaspectratio=true,clip=true]{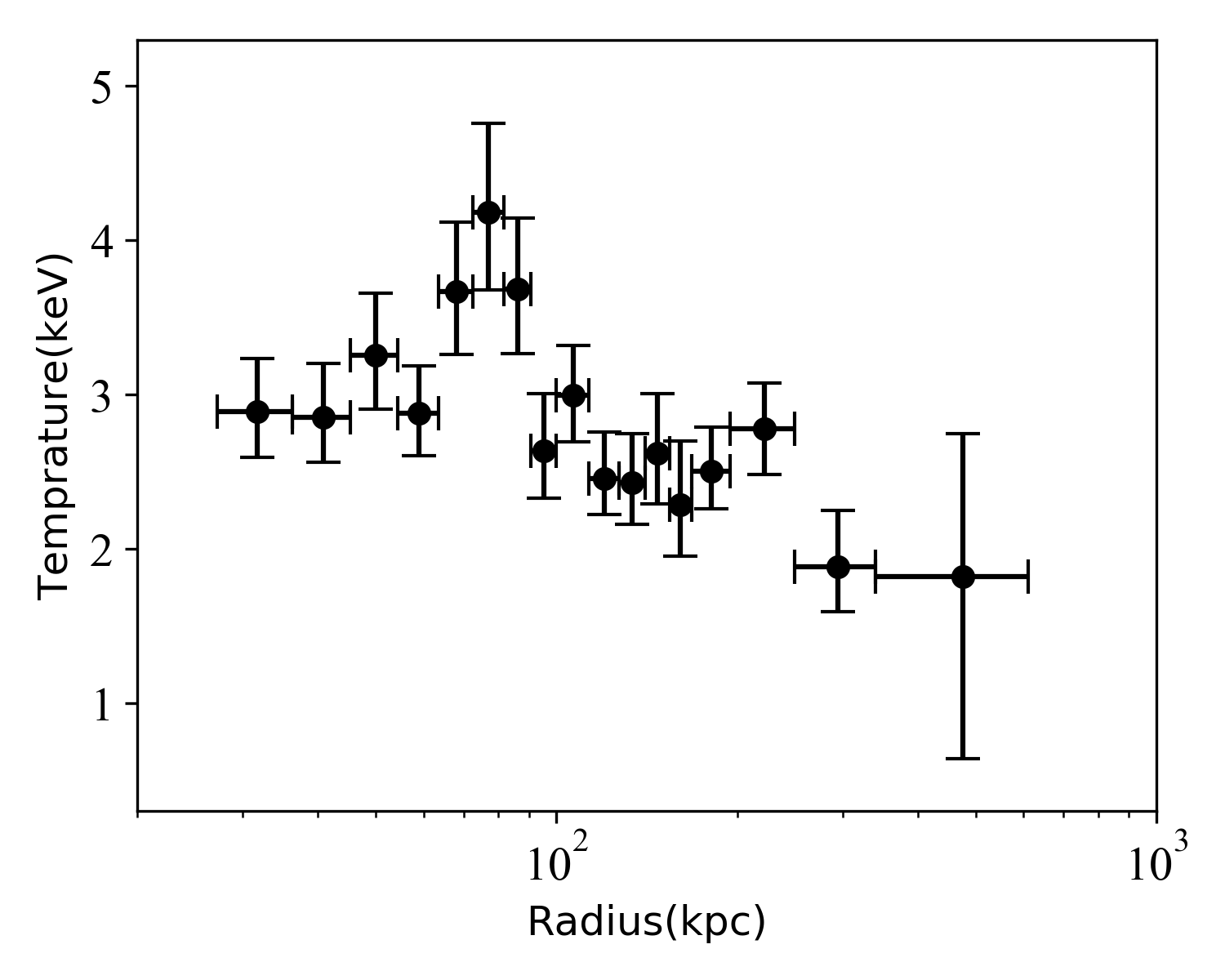}
        \caption{ Fitting results of radial annular regions. Left-hand panel: \cha\ 0.5-2 keV image overlaid with \gmrt\ 610 MHz contours. The \gmrt\ observation code is 21\_066 (PI: Joydeep Bagchi) with $5''$ FWHM resolution. 
        The X-ray image is binned with a size of two 0.492$''$ pixels and smoothed with a uniform Gaussian kernel of 1.97$’'$.
        The radius of the rings gradually increases from 30$''$ to 675$''$ i.e. 27.2 to 612.2 kpc. Right-hand panel: the temperature as a function of radius shows that A407 has a weak cool core and the gas temperature reaches a maximum, $4.18^{+0.58}_{-0.50}$ keV, at 85 kpc. The temperature and abundance values are given in Table~\ref{tab:ring_label}.
        }
        \label{pic rings result}
    \end{figure*}

A global spectrum, extracted excluding the central region, gives a mean X-ray temperature of 2.7 keV. This implies a cluster total mass of ${M_{500}} = 1.4 \times 10^{14} {M_{\odot}}$ and a radius of $r_{500}=774$ kpc according to the $M_{500}-T_{500}$ relation from \cite{2009ApJ...693.1142S}.
Here $r_{500}$ refers to the radii within which the cluster mass density is 500 times the universe critical density and $M_{500}$ is the mass within $r_{500}$.
We also extract the spectrum from the annuli centered on the X-ray peak to produce the temperature profile as shown in Fig.~\ref{pic rings result}.
The region of the central galaxies is excluded to avoid their contamination. The radii of the annuli gradually increase from 30$''$ to 675$''$. The radius is increased by steps of 10$''$ from 30$''$ to 110$''$ and 15$''$ from 110$''$ to 185$''$. For the outer 4 large regions, their radii are increased by 30$''$, 60$''$, 100$''$, and 300$''$ respectively. 
As detailed in Section~\ref{sec:data reduction}, we use the double subtraction method to fit the cluster's spectrum. The energy range is chosen as 0.5-8.0 keV. The best-fitting results are shown in Fig.~\ref{pic rings result}. The temperature profile indicates that A407 has a cool core (CC) and the gas temperature reaches a maximum, $4.18^{+0.58}_{-0.50}$ keV, at 85 kpc. 
We also notice that the temperature rises at $\sim$ 250 kpc, which implies there may exist hot regions.

    %panda7 reg 

Alternatively, the spectra from sector regions are fitted to analyze the interaction of radio jets and cluster hot gas. The inner radius of the sector regions is 30$''$ and the outer radius is 115$''$. The fitting results are shown in Fig.~\ref{pic panda result}. The regions that contain the radio jets (region 1 and 4) have the maximum temperatures which are $4.00_{-0.36}^{+0.36}$ and $3.57_{-0.41}^{+0.45}$ keV respectively. Region 1 and 7 have the highest abundances, $0.99_{-0.30}^{+0.39}$, and $0.68_{-0.21}^{+0.28}$, which indicates that the higher abundance value may be produced as the jets uplift high abundance gas from the galaxy core. 

\begin{figure}
        \centering
        \includegraphics[width=0.47\textwidth,keepaspectratio=true,clip=true,]{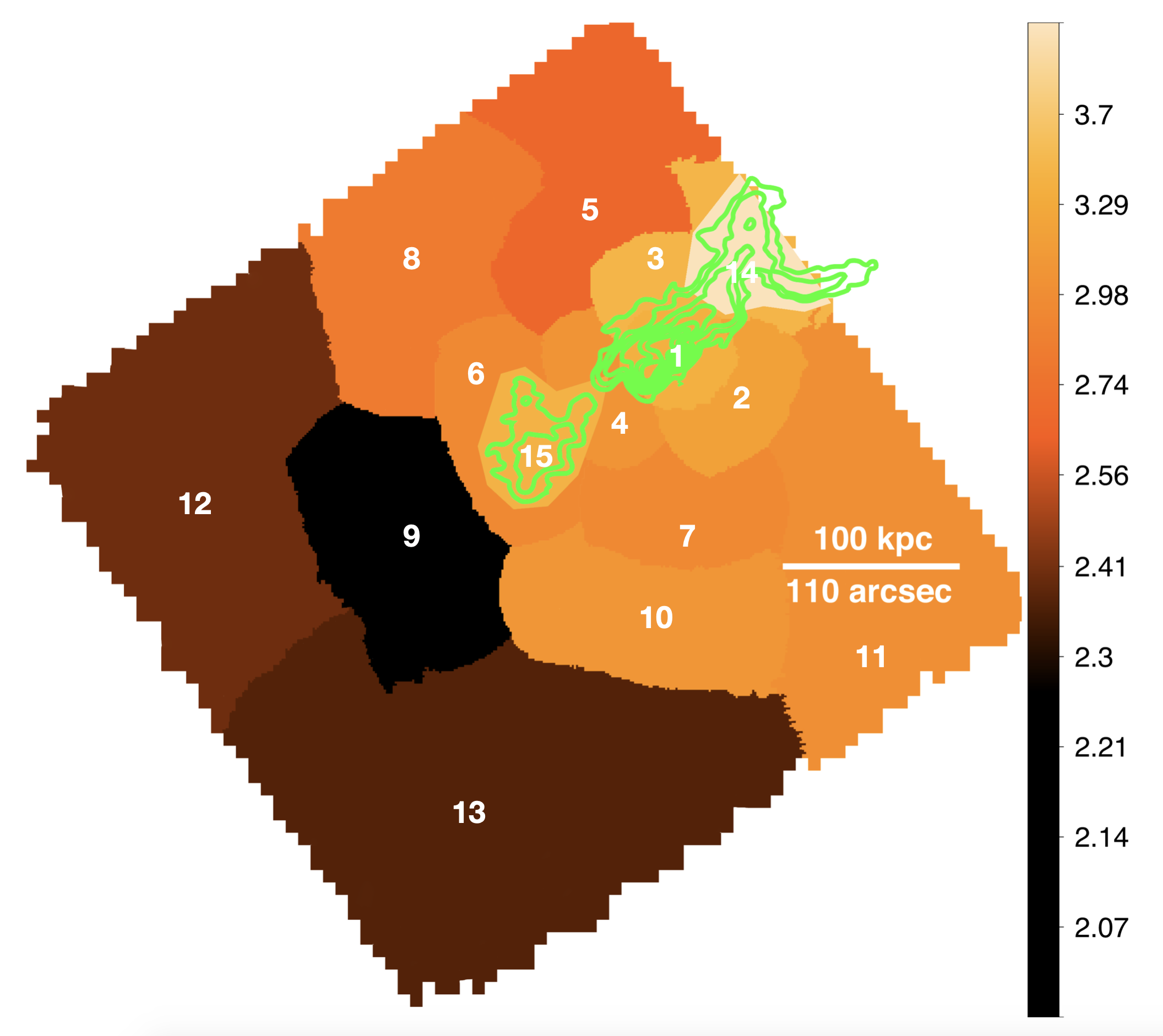}
        \caption{The temperature map generated by \textit{Contour Binning} and overlaid with \gmrt\ 610 MHz contours. The temperature values are listed in Table~\ref{tab:contourbin_label}. 
        The white numbers marked on the picture are the region identifiers. Regions 14 and 15, where the radio jets lie, have relatively higher temperature than others. 
        }
        \label{pic temperature map}        
\end{figure}

To analyze the gas structure of the cluster, we derive the ICM temperature map. We extract spectra from ACIS-S3 where the cluster mainly fits. 
We generate adaptive spatial bins to fit the cluster temperature using `Contour binning' \citep{2006MNRAS.371..829S}, an algorithm for binning X-ray data by following the surface brightness of the X-ray image. 
After masking the point sources of the exposure corrected X-ray image, we set the signal-to-noise ratio to 30, which requires $\sim900$ background-subtracted counts per bin. We set the geometric constraint value to 1.3, ensuring the shape of bins is not too elongated. 
As shown in Fig.~\ref{pic temperature map}, we use these regions to extract the spectra and response files and fit the gas temperature. We use two regions (region 14 and 15) to cover the radio jet regions and add the fitting result in the temperature map. The detailed temperature of each region are listed in Table~\ref{tab:contourbin_label}.
We note that regions 14 and 15, where the radio jets lie, have relatively higher temperatures than others.

\begin{figure}
        \centering
        \includegraphics[width=0.47\textwidth]{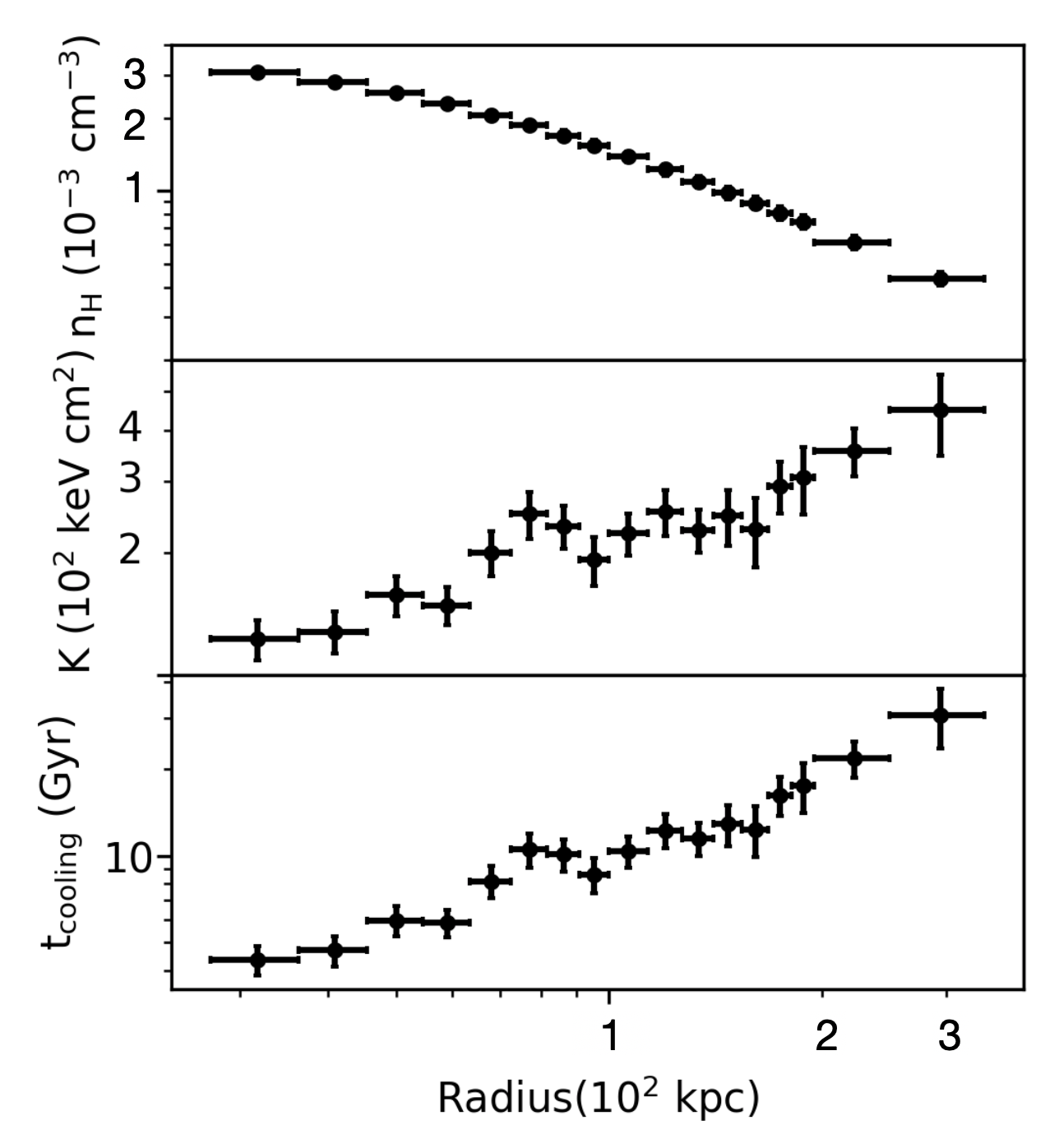}
        \caption{Top panel: Hydrogen density profile of A407. Middle panel: Entropy profile of A407. Bottom panel: cooling time profile of A407. The central region has an entropy of 108 $\rm{keV\ cm^2}$ and a cooling time of 4.4 Gyr. See Section~\ref{sub:4.2hotgas} for details.}
        \label{pic entropy}
    \end{figure}

Entropy is also a useful quantity to study the thermal gas history of a cluster as it reflects ICM global properties. It is defined as $K=T_{X} n_{e}^{-2 / 3}$ \citep{1999Natur.397..135P,2000MNRAS.315..689L}, where $T_{X}$ is X-ray temperature and $n_{e}$ is electron density.
With the SBP and temperature profile, we can calculate the entropy profile of A407. 
Assuming spherical symmetry, the cluster's deprojected hydrogen density profile as a function of the physical off-center radius is (e.g. \citealt{1976A&A....49..137C,1988xrec.book.....S}): 

\begin{equation} \label{eq:betamodel}
\centering
    n_{\mathrm{H}}=n_{0}\left(1+\frac{r^{2}}{r_{\mathrm{c}}^{2}}\right)^{-\frac{3}{2} \beta}
\end{equation}
where $n_{0}$ represents the central density and we use the equation (4) of \cite{2018MNRAS.481.4111G} to derive it:

\begin{equation}
\centering
    n_{0}=\frac{3600 \times 180}{\pi} \sqrt{\frac{10^{14} 4 \sqrt{\pi} I_{0} \Gamma(3 \beta)}{\left(\frac{n_{e}}{n_{H}}\right) \frac{f_{x}}{\eta} r_{c} \Gamma(3 \beta-1 / 2)}}
\end{equation}
This equation uses parameters ($I_{0}, \beta, r_{c}$) listed in Table~\ref{tab:beta_label} from the $\beta$-model fit of the cluster image. $n_{e}$ and $n_{H}$ are the electron and hydrogen number densities and $n_{e} \simeq 1.2 n_{H}$. $\Gamma$ is the gamma function. $\frac{f_{x}}{\eta}$ is the ratio between the X-ray photon flux and the normalization of the {\tt APEC} model. With the density profile and the temperature map, A407's entropy profile is generated and shown in the middle panel of Fig.~\ref{pic entropy}. 
We also calculate the central entropy of A407, which is 108 $\mathrm{keV} \mathrm{cm}^2$.

As detailed in \cite{2010A&A...513A..37H}, from the density profile and temperature profile, we can estimate the average cooling time profile of the gas using:

\begin{equation}
    t_{\mathrm{cool}}=\frac{3}{2} \frac{\left(n_{\mathrm{e}}+n_{\mathrm{i}}\right) k T}{n_{\mathrm{e}} n_{\mathrm{H}} \Lambda(T, Z)}
\end{equation}

where $n_{\mathrm{i}}$ and $n_{\mathrm{e}}$ are the number density of ions and electrons respectively. $\Lambda(T, Z)$ is determined from the cooling curve in \cite{2009A&A...508..751S}. The cooling time profile is shown in the bottom panel in Fig.~\ref{pic entropy}. From the cooling time profile, we note that the cooling time of the central region is 4.4 Gyr.

\cite{2010A&A...513A..37H} test several parameters to distinguish cool core (CC), weak CC and non-cool-core (NCC) clusters. 
They find that the central cooling time (measured at 0.048 $r_{500}$) is the best parameter to distinguish among the classes for low redshift clusters. 
Weak CC are characterized as having central cooling times between 1 and 7.7 Gyr. They also suggest the entropy of weak CC clusters is between $\sim22$ and $\sim150$ $\mathrm{keV} \mathrm{cm}^2$. 

As the central cooling time of A407 is 4.4 Gyr and the central entropy is 108 $\mathrm{keV} \mathrm{cm}^2$, A407 is a typical weak CC cluster. The weak CC size of A407 is $\sim60$ kpc where the cooling time equals the time from $z = 1$ to $z = 0$.

\section{Discussion} %(fold)
\label{sec:discussion}

\subsection{Merging activity of central galaxies}
\label{subsec: merging activity}
Two broad classes of physical processes are responsible for the interaction of galaxies in high-density environments. 
The first class includes the gravitational processes which include all sorts of tidal interactions. 
The second class includes the hydrodynamic interactions between the galaxy ISM and the hot ICM, such as viscous stripping, ram pressure stripping, and thermal evaporation \citep{2006PASP..118..517B}. 

Previous observations of A407 indicate that it possibly provides unique evidence for the formation of a BCG by mergers. 
\cite{1982ApJ...263...14S} suggest this system is a forming cD by merging events as it harbors a remarkably compact group of galaxies and has an extended stellar envelope. 
\cite{2017MNRAS.471..617B} present the \gmrt\ low-frequency radio images at 610, 235, and 150 MHz and they find a very peculiar twisted jet system which may be caused by the galaxy merging activities. 
\cite{2015A&A...579A..27S} reveal the two episodes of AGN activity with the \lofar\ data. The jet precession implies that the AGN is intermittently active as it moves in the dense cluster core. 
\cite{2020MNRAS.491.2617E} investigate the integral spectroscopy of 23 BCGs and determine the age, metallicity, velocity, and velocity dispersion of stars in each. Their results are consistent with the scenario that BCG cores formed long ago (with an average age of $13.3\pm2.8$ Gyr) and the outer intracluster light (ICL) formed more recently. However, they also notice some local BCGs are still assembling and A407 is one such example.

\cha's high spatial resolution data reveal more information about the merging process of A407. 
If it is a rapid merging, we would see signs of gas stripping or no X-ray coronae at all. On the other hand, if the galaxies are in a process of a slow `merger of equals', we might expect to see X-ray coronae associated with several of the bright galaxies, following the $L_{Ks}$-$L_{X}$ relation \citep{2007ApJ...657..197S}. 
As detailed in Section~\ref{sec:results}, seven out of nine optical bright galaxies are detected in the soft X-ray band. However, the detection limit of \cha\ with an exposure time of 43 ks is $\sim 10^{40}\ \rm{erg\ s^{-1}}$. The $L_{X}$ of galaxies are close to this limit. 
Therefore, the two remaining galaxies, G6 and G8, may also have faint X-ray corona which could be confirmed with longer exposures.

One possible explanation for the existence of the galaxy corona is that the galaxies are in the process of a `slow merge' which doesn't strip all the galaxy gas halos. 
Another explanation is that the galaxies with coronae are far from each other along the line of sight. This leads to the survival of the coronae while some other galaxies, such as G6 and G8, likely lost their halos in the merging progress. 
However, considering that the velocities of the galaxies (Table~\ref{table galaxies}) vary little, the `slow merge' model is preferred in A407. 
We also calculate the $L_{X}$ of each galaxy based on the $L_{Ks}$-$L_{X}$ relation. 
However, the merging activities may produce the diffuse stellar halos of each galaxy. 
Therefore, we scale the $L_{Ks}$ of each galaxy with a value of 2.35, which is the ratio of the total $Ks$ band luminosity inside the radius of 35$''$ and the sum of $L_{Ks}$ of the central nine galaxies.
The predicted $L_{X}$ values from $L_{Ks}$ are shown in the last column of Table~\ref{table galaxies}, which are close to the observed $L_{\rm 0.5 - 2 keV}$ listed in Table~\ref{table galaxies}. 
Despite the large scatter in the $L_{Ks}$-$L_{X}$ relation, this result supports the `slow merge' model that coronae of the galaxies are not disrupted entirely in the merging progress.

We calculate the required mass to bind the central galaxies gravitationally with the equation:

\begin{equation}
    \label{equation: binding mass}
    M=\frac{3 R \sigma^{2}}{G}
\end{equation}

where $\sigma$ is the line of sight velocity dispersion of central galaxies in the rest-frame and $\sigma = 584$ ${\rm km\ s^{-1}}$. The radius R is 35$''$ (31.7 kpc) which contains all nine galaxies and their extended emission. %According to this equation, 
The required binding mass is $M_{\rm bind}=7.5 \times 10^{12} M_{\odot}$. 
As a comparison, we remove galaxy G4, which has a most different velocity from those of other central galaxies. The value of $\sigma$ reduces to 413 ${\rm km\ s^{-1}}$ and the binding mass reduces to $M_{\rm bind}=3.8 \times 10^{12} M_{\odot}$. 

We also calculate the total mass within 35$''$ with the dynamical mass-to-light ($M/L$) relation, which includes the dark matter.  \cite{2006MNRAS.366.1126C} investigated the dynamical $M/L$ ratio of elliptical galaxies. They fitted the correlation between the $I$-band $M/L$ and the $Ks$-band luminosity:

\begin{equation}
    (M / L_I)=(1.88 \pm 0.20)\left(\frac{L_{K}}{10^{10} L_{K, \odot}}\right)^{0.32 \pm 0.05}
\end{equation}

The $I$-band luminosity within 35$''$ of A407 is measured from SDSS data ($L_I = 3.1\times10^{11} L_{\odot}$) and the $Ks$-band luminosity is from the \twomass\ data ($L_{ks} = 8.5\times10^{11} L_{\odot}$). Thus, the dynamical mass within 35$''$ is $M_{\rm d} = 2.4 \times 10^{12} M_{\odot}$. This value is close to the result that calculated by \cite{2017MNRAS.471..617B}. They calculated the dynamical mass within the radius of 30$''$ as $2.2 \times 10^{12} M_{\odot}$ from the halo radial velocity dispersion. Thus, the central dynamical mass of A407 is close to the binding mass calculated above. 

Many early-type galaxies (ETGs) have distinctive tidal tails. These tidal tails may be produced during the merging process \citep{1983ApJ...274..534M,2005AJ....130.2647V,2008ApJ...677..846B,2010ApJ...715..972J}. We might also expect such features in the optical band of A407. However, there are no stellar trails detected in the \sdss\ $r$ band image in Fig.~\ref{pic overview}(c). Deeper optical observations will help to reveal more details of the merger by detecting such features.

\subsection{The future BCG } % (fold)
\label{sub:bcg_mass}
One interesting question is, when the central galaxies merge, what will be the mass of the central BCG? %First, we derive the BCG's stellar mass. 
The stellar mass of a galaxy can be calculated from the stellar $M/L$ ratio relation \citep{2003ApJS..149..289B}. We choose a circle with a radius of 35$''$, which contains all nine galaxies and their extended emission. From the \twomass\ $K_s$ band image, the total $K_s$ band luminosity within this circle is $(8.5\pm0.5)\times10^{11} L_{\odot}$. Thus, the corresponding stellar mass of the future BCG is $(7.4\pm0.4)\times10^{11} M_{\odot}$ using the stellar $M/L$ ratio relation \citep{2003ApJS..149..289B}. 

There is a correlation between BCG stellar mass and the host cluster mass. It can be fit with a power-law of the form $M_{\mathrm{BCG}}=\beta M_{\mathrm{Cluster}}^{\alpha}$ where $M_{\mathrm{BCG}}$ is the BCG stellar mass and $M_{\mathrm{Cluster}}$ is $M_{200}$ which represents the cluster mass within $r_{200}$ \citep{2012MNRAS.427..550L,2016MNRAS.460.2862B}. 
Accordingly, we can check whether A407 and its BCG follow this mass relation. The cluster mass $M_{500}=1.4 \times 10^{14} {M_{\odot}}$ (from the $M_{500}-T_{500}$ relation; \cite{2009ApJ...693.1142S}). We then convert ${M_{500}}$ to ${M_{200}}$ by assuming the cluster follows an NFW mass profile \citep{1997ApJ...490..493N} which gives $r_{500} / r_{200}=0.669$. The resulting ${M_{200}}$ of A407 is $1.9 \times 10^{14} {M_{\odot}}$. We find the masses of A407 and its future BCG are consistent with this relation. Thus the predicted value of the mass of the future BCG is reasonable.

To estimate the merging time of this system, we use the mean merging time-scale relation proposed by \cite{2008MNRAS.391.1489K}. They investigate the major merger rates based on the Millennium N-body simulation and calibrate the results from deep galaxy surveys. Their work shows that, for samples with stellar masses $\mathrm{M}_{*}>5 \times 10^{9} {M}_{\odot}$ and radial velocity differences $\Delta v<3000 \mathrm{~km} \mathrm{~s}^{-1}$, the characteristic mean merging time-scale of galaxy pairs in the low-redshift region ($z \leq 1$) could be fitted well by:

\begin{equation}
    \left\langle T_{\text {merge }}\right\rangle=3.2 \mathrm{Gyr} \frac{r_{\mathrm{p}}}{50 \mathrm{kpc}}\left(\frac{M_{*}}{4 \times 10^{10} h^{-1} {M}_{\odot}}\right)^{-0.3}\left(1+\frac{z}{20}\right)
    \label{merge time scale}
\end{equation}
$\left\langle T_{\text {merge }}\right\rangle$, $r_{\mathrm{p}}$ and ${M}_{*}$ represent the mean merging time of galaxy pairs, projected separation (kpc), and stellar mass of the galaxy pairs (${M}_{\odot}$). 
The projected distance of A407's central galaxies can be obtained from the optical image (Fig.~\ref{pic overview}). The masses of galaxies are obtained from their $L_{Ks}$ which are derived from the \twomass\ point source catalog (PSC) (Table~\ref{table galaxies}). 
However, galaxies are extended in the \twomass\ optical image. Thus, these $L_{Ks}$ are underestimated. 
As detailed in Section~\ref{subsec: merging activity}, we apply a scale value of 2.35 to each galaxy to derive $L_{Ks}$ and stellar mass.
We obtain the total mass of each galaxy pair and calculate the merging time-scale by equation~\ref{merge time scale}. 
The merging time-scale range is $0.3 \sim 2.3$ Gyr which provides a rough estimate of the central merging time of A407.
We note that the estimation is based on the two body merging time-scale, the merging process of this compact group is much more complicated and the final merging time is likely larger.

\subsection{Host of the Radio AGN 4C~35.06}
\label{Host of the Radio AGN}
Which galaxy is the host of 4C~35.06?
The {\em VLBA} observations by \cite{2010A&A...516A...1L} show pc-scale radio emission centered at the galaxy G3, which suggests that G3 is radio active now.
\cite{2015A&A...579A..27S} derived the radio luminosity of 4C 35.06 to be $L_{1.4\ \mathrm{GHz}} = 2.5 \times 10^{24}\, \mathrm{W}\, \mathrm{Hz}^{-1}$, which makes it a strong radio AGN. 
They suggested that the rapid movement of galaxy G3 and its episodic AGN radio activities are the reasons for the observed peculiar radio morphology. 
They also suggest that G3 switched off its radio emission and then restarted while it was moving to its current position, which leads to the offset inner double-lobed morphology.
\cite{2015A&A...579A..27S} also detected a broad HI absorption feature (FWHM of 288 ${\rm km\ s^{-1}}$) on top of the strong radio core of 4C~35.06 at a velocity of $\sim$ 13858 ${\rm km\ s^{-1}}$. Assuming the HI gas has a similar velocity to that of the host galaxy for 4C~35.06, only G2, G3, and G7 have velocities within 300 ${\rm km\ s^{-1}}$ of that of the HI absorption feature, as shown in Table~\ref{table galaxies}.
Moreover, strong radio AGN are typically associated with massive galaxies. As listed in Table~\ref{table galaxies}, G1, G3, and G7 have the largest galaxy masses. Thus, G3 would be the most likely host from these constraints, as G7 seems to be too far away from the radio center.
However, \cite{2017MNRAS.471..617B} suggested that the host of the radio AGN is G6 which lies at the center of the principal jet direction. Though G6 is faint in both optical and infrared light, they suggest it may have been stripped of the majority of its outer halo stars in multiple tidal encounters, while still retaining a dense stellar core at the center. 

With the \cha\ data, we could reveal more information of the central galaxies. \cite{2005MNRAS.362...25B} and \cite{2007MNRAS.376.1849H} suggest that radio AGN activity is associated with the cooling of the hot gas from the halos of elliptical galaxies and clusters. \cite{2009ApJ...704.1586S} presented a systematic analysis of 152 groups and clusters, and claimed that every BCG with a strong radio AGN ($L_{1.4\ \mathrm{GHz}}>2 \times 10^{23}\, \mathrm{W}\, \mathrm{Hz}^{-1}$ ) has an X-ray CC either in the large cluster CC class or the galaxy corona class. 
Since the radio luminosity of 4C 35.06 is $L_{1.4\ \mathrm{GHz}} = 2.5 \times 10^{24}\, \mathrm{W}\, \mathrm{Hz}^{-1}$, which is larger than $2 \times 10^{23}\, \mathrm{W}\, \mathrm{Hz}^{-1}$, it should have an X-ray CC. 
As mentioned in Section~\ref{sec:results}, A407 is a weak CC cluster and G3 has the most prominent X-ray corona and lies closer to the radio core. 
Therefore, we suggest G3 is the most likely host of the luminous radio AGN. Of course, an ambiguous determination of the host would require radio images at higher angular resolution. As G5 hosts the brightest X-ray AGN in this group and likely hosts a massive SMBH from the analysis by \cite{2017MNRAS.471..617B}, we cannot exclude the possibility that G5 also contributes to the radio emission of 4C~35.06.

    \begin{figure}
        \centering
        \includegraphics[width=0.47\textwidth]{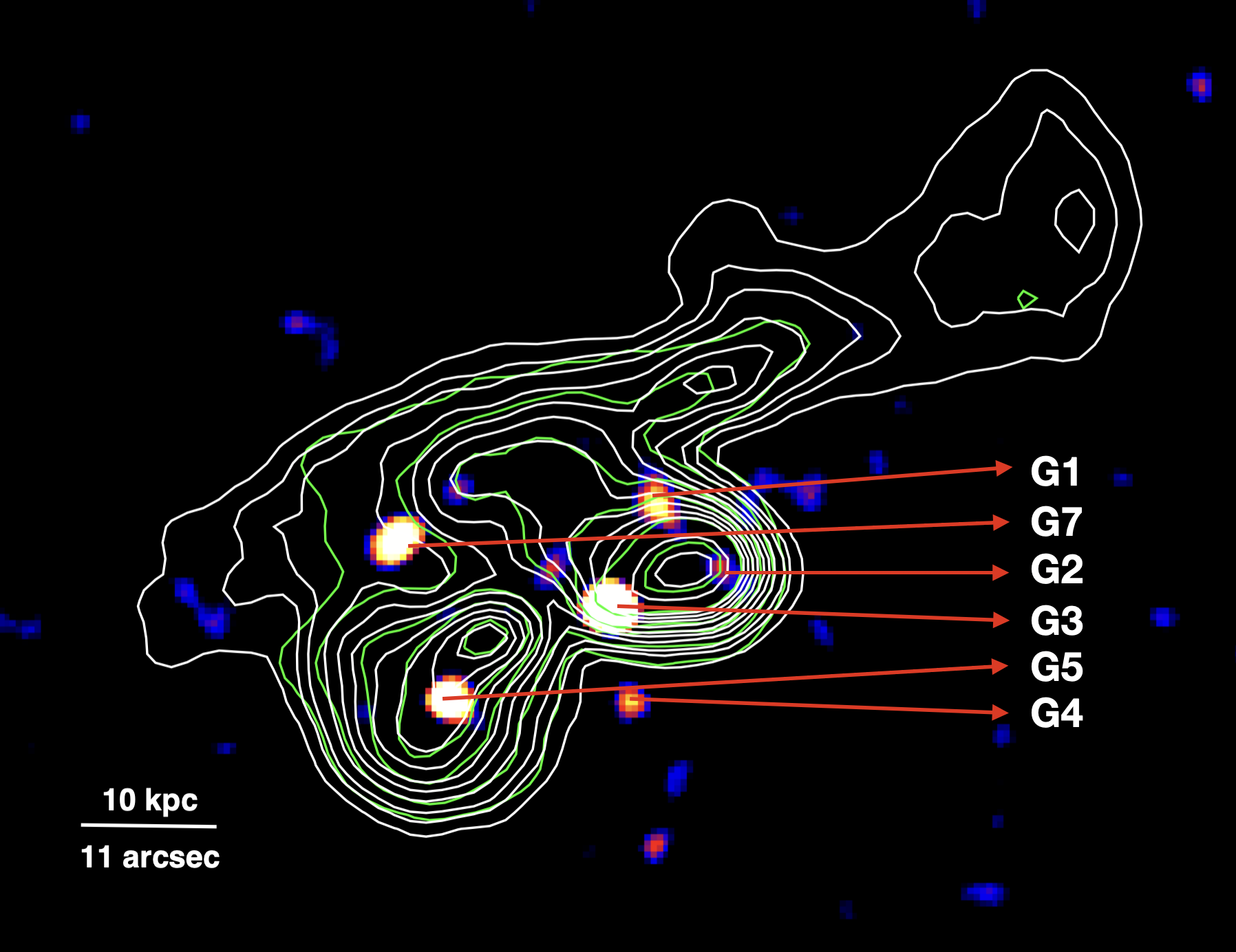}
        \caption{\cha\ 0.5-2 keV image overlaid with \vla\ 1.5 GHz (white) and 4.9 GHz (green) contours.}
        \label{pic radio contour}
    \end{figure}

\subsection{Feedback from AGN} % (fold) 
AGN can release tremendous amounts of energy to their surroundings.  
\cite{1993MNRAS.264L..25B} noted the existence of radio-filled cavities in NGC1275 based on ROSAT observations.
\cite{2000ApJ...534L.135M} found a close association between the radio lobes and X-ray cavities in Hydra A which could suggest the gas was displaced as the radio lobes expanded. Thus X-ray cavities can provide a direct way to measure the energy injected by AGN \citep{2000A&A...356..788C,2001ApJ...554..261C}.

The energy from radio jets are near those required to offset radiative cooling \citep{2012ARA&A..50..455F,2012NJPh...14e5023M}.\cite{2009AIPC.1201..198N} studied a sample of nearby giant elliptical galaxies from {\em Chandra} observations. Their results show that the average power injected by AGN is sufficient to stop the hot atmospheres from cooling. 
Not only the temperature, heavy elements can also be transported from the core by the large-scale outflows and change the ICM abundance \citep{2007ARA&A..45..117M,2008A&A...482...97S,2011ApJ...731L..23K,2015MNRAS.452.4361K}. 

\cite{2017MNRAS.471..617B} suggested the possibility that radio jets in this source might be heating and interacting with the ICM via a jet-mode feedback. The temperature map of A407 does indicate the presence of feedback features including heating and abundance enrich. As shown in Fig.~\ref{pic panda result}, the regions that overlap with the radio jets (region 1 and 4) have the maximum temperatures i.e. $4.0_{-0.4}^{+0.4}$ and $3.6_{-0.4}^{+0.5}$ keV. This implies that the gas around the radio jets is heated by the feedback. As can be seen from Table~\ref{tab:panda_label}, region 1 and 7, which include the north-west radio jet, have the highest abundances, $1.0_{-0.3}^{+0.4}$, and $0.7_{-0.2}^{+0.3}$ respectively.  The increase in temperature and abundance along the jet direction is also seen in Fig.~\ref{pic temperature map} and Table~\ref{tab:contourbin_label}. 

The power of cavities can provide a lower limit to the jet energy as it is a direct evidence for AGN heating \citep{2008ApJ...686..859B}. 
\cite{2008ApJ...686..859B}, \cite{2010ApJ...720.1066C}, and \cite{2011ApJ...735...11O} investigate the scaling relationship between mechanical jet power and radio luminosity. 
Based on their work, we estimate the jet power's lower limit in A407 from its 1.4 GHz radio luminosity ($L_{1.4\ \mathrm{GHz}}=2.5 \times 10^{24}\, \mathrm{W}\, \mathrm{Hz}^{-1}$ ). 
The lower limit of the mechanical jet power is $1.0 \times 10^{44}\,\mathrm{erg}\, \mathrm{s}^{-1}$. 
 
As detailed in Section~\ref{sub:4.2hotgas}, no cavity structure has been found in A407 yet. It may need deeper X-ray observation to confirm. Assuming there exist two cavities with the same scale of the 610 MHz radio contour in Fig.~\ref{pic panda result} and their shape could be approximated by spheres with the radius of 40 kpc, we can estimate the mechanical pV energy of the cavities based on the gas temperature and density profile \citep{2012NJPh...14e5023M}. The total mechanical energy of the two cavities is $5.8\times 10^{59}$ erg. As the mechanical jet power is $1.0 \times 10^{44}\,\mathrm{erg}\, \mathrm{s}^{-1}$, it would take 0.18 Gyr for the radio AGN to supply the cavities' mechanical pV energy. For comparison, the central cooling time of A407 is 4.4 Gyr.

Entropy can also reflect ICM global properties. \cite{2009ApJS..182...12C} present radial entropy profiles for 239 clusters from the {\em Chandra} X-ray Observatory Data Archive. 
They find that most entropy profiles are well fitted by a power law at large radii and a constant value at small radii: $K(r)=K_{0}+K_{100}(r / 100 \mathrm{kpc})^{\alpha}$. $K_0$ is the typical excess of core entropy above the best-fitting power law found at larger radii. For clusters with temperature $T_{X}<4\, \mathrm{keV}$, the $K_0$ distribution peaks at $K_{0} \sim 15\, \mathrm{keV}\,\mathrm{cm}^{2}$. 
For clusters with temperature $T_{X}>8\, \mathrm{keV}$, the $K_0$ distribution peaks at $K_{0} \sim 150\, \mathrm{keV}\,\mathrm{cm}^{2}$. 
The mean temperature of A407 is 2.7 keV and the entropy profile is shown in Fig.~\ref{pic entropy}. From the figure, we notice that the entropy profile of A407 is relatively higher compared to other clusters with temperatures lower than $T_{X}<4$ keV in \cite{2009ApJS..182...12C}. The entropy at small radii is $\sim 200\, \mathrm{keV}\,\mathrm{cm}^{2}$. The higher entropy in A407 at small radii might be caused by the AGN feedback. 

In summary, A407 shows evidence that its ICM temperature, abundance, and entropy are affected by AGN feedback, most likely from the radio jets.

\section{summary} 
\label{sec:summary} 
A407 contains a compact group of galaxies in its central region. We study A407 with the \cha\ observation combined with previous radio and optical observations. Our major results and conclusions are: 
 
\begin{itemize} 
\item A407 has a weak CC within 60 kpc. The central region of A407 has a cooling time of 4.4 Gyr and an entropy of 108 $\mathrm{keV} \mathrm{cm}^2$. The ${M_{200}}$ of A407 is $1.9 \times 10^{14}\, {M_{\odot}}$. 
 
\item Due to the existence of the central galaxy X-ray coronae, we suggest the central galaxies are in the process of a `slow merge’. When these galaxies merge, we predict the future BCG's stellar mass would be $7.4\times10^{11} {M_{\odot}}$ and the merging time-scale range is $0.3 \sim 2.3$ Gyr.  The predicted BCG mass follows the mass relation between BCGs and host clusters (e.g. \citealt{2012MNRAS.427..550L,2016MNRAS.460.2862B}). 
 
\item For galaxies in the central region of A407, 
G3 has the most prominent X-ray thermal corona. Combining the results of X-ray and radio observations, we suggest galaxy G3 is the host of radio AGN 4C 35.06. 

\item We also find evidence of AGN feedback in the hot gas of A407. The regions that overlap with the radio jets have higher temperatures and abundance than other regions. The entropy of A407's center is also relatively higher than other clusters which may be due to the effects of AGN feedback.
\end{itemize}

In summary, A407 has a very unusual cluster core in optical, which is populated by a cluster of galaxies instead of a single dominant galaxy as in most clusters.
We witness it in a crucial formation epoch accompanied by the helically twisted jets and merging features which have the potential of clarifying how BCGs originate.  We present a study focused on the X-ray band to reveal merging details of the central galaxies and implications of the AGN feedback activities. A407 suggests the scenario in which AGN activity can be triggered during the formation of the cluster BCG. Deeper X-ray observations would reveal more details of the merging and AGN feedback process. Further optical observation would help to confirm if there exist any stellar tidal trails. More cases, from both low and high redshifts, are needed to reveal more detail about the BCG formation processes.

\section*{Acknowledgements}
We thank the referee, Dr. Joydeep Bagchi, for comments.
The scientific results reported in this article are based on the observation made by the \cha\ X-ray Observatory, ObsIDs 18267, which is operated by the Smithsonian Astrophysical Observatory (SAO) for and on behalf of the National Aeronautics Space Administration (NASA). This publication makes use of data products from the Two Micron All Sky Survey, which is a joint project of the University of Massachusetts and the Infrared Processing and Analysis Center/California Institute of Technology, funded by the National Aeronautics and Space Administration and the National Science Foundation. We also present data obtained with the Giant Meter-wave Radio Telescope (\gmrt), which is run by the National Centre for Radio Astrophysics of the Tata Institute of Fundamental Research. We use the images from \sdss\ and the funding for \sdss\ is provided by Alfred P. Sloan Foundation, the U.S. Department of Energy Office of Science, and the Participating Institutions. We also made use of the NASA/IPAC Extragalactic Database (NED) which is operated by the Jet Propulsion Laboratory, California Institute of Technology, under contract with NASA. 
D.V.L. acknowledges the support of the Department of Atomic Energy, Government of India, under project no.12-R\&D-TFR-5.02-0700.
L.J. is supported by NSFC grant U1531248 and BICCAS grant 114332KYSB20180013.
W.F. and C.J. acknowledge support from the Smithsonian Institution and the \cha\ High Resolution Camera Project through NASA contract NAS8-03060.  

\section*{DATA AVAILABILITY}

The \cha\ raw data used in this paper are available to download at the \cha\ Data Archive\footnote{https://cxc.harvard.edu/cda/} and HEASARC Data Archive\footnote{https://heasarc.gsfc.nasa.gov/docs/archive.html} website.
The reduced data underlying this paper will be shared on reasonable requests to the corresponding authors.

    \bibliographystyle{mnras}
    \bibliography{citations}{}

\begin{table*}
\centering
\caption{The fitting results of the radial annular regions.}
\resizebox{\textwidth}{!}{
\begin{tabular}{lcccccccccc}
\hline\hline 
Radius (arcsec) & 30-40 & 40-50 & 50-60 & 60-70 & 70-80 & 80-90 & 90-100 & 100-110 & 110-125  \\
Radius (kpc) & 27.2-36.2 & 36.2-45.3 & 45.3-54.4 & 54.4-63.4 & 63.4-72.5 & 72.5-81.6 & 81.6-90.7 & 90.7-99.7 & 99.7-113.3 \\
Temperature (keV) & $2.89_{-0.30}^{+0.34}$ & $2.85_{-0.29}^{+0.35}$ & $3.25_{-0.35}^{+0.40}$ & $2.88_{-0.27}^{+0.31}$ & $3.67_{-0.41}^{+0.45}$ & $4.18_{-0.50}^{+0.58}$ & $3.68_{-0.42}^{+0.46}$ & $2.64_{-0.31}^{+0.37}$ & $2.99_{-0.30}^{+0.33}$ \\
Abundance & $0.32_{-0.16}^{+0.23}$ & $0.31_{-0.15}^{+0.21}$ & $0.18_{-0.14}^{+0.19}$ & $0.47_{-0.19}^{+0.25}$ & $0.60_{-0.26}^{+0.35}$ & $0.76_{-0.32}^{+0.46}$ & $0.67_{-0.28}^{+0.38}$ & $0.17_{-0.13}^{+0.19}$ & $0.50_{-0.19}^{+0.25}$  \\
\hline
Radius (arcsec) & 125-140 & 140-155 & 155-170 & 170-185 & 185-215 & 215-275 & 275-375 & 375-675 & \\
Radius (kpc) & 113.3-126.9 & 126.9-140.5 & 140.5-154.1 & 154.1-167.7 & 167.7-195.1 & 195.0-249.4 & 249.4-340.1 & 340.1-612.2 & \\
Temperature (keV) & $2.46_{-0.23}^{+0.30}$ & $2.43_{-0.27}^{+0.32}$ & $2.62_{-0.33}^{+0.39}$ & $2.29_{-0.33}^{+0.42}$ & $2.51_{-0.25}^{+0.28}$ & $2.78_{-0.29}^{+0.30}$ & $1.89_{-0.29}^{+0.36}$ & $1.82_{-1.18}^{+0.92}$ & \\
Abundance & $0.18_{-0.11}^{+0.14}$ & $0.54_{-0.23}^{+0.34}$ & $0.18_{-0.13}^{+0.19}$ & $0.13_{-0.11}^{+0.19}$ & $0.26_{-0.12}^{+0.16}$ & $0.53_{-0.17}^{+0.22}$ & $0.24_{-0.18}^{+0.30}$ & $0.23_{-0.23}^{+0.90}$ & \\
\hline
\end{tabular}
}
\label{tab:ring_label}
\end{table*}

\begin{table*}
    \centering
    \caption{The fitting results of the 7 panda shape regions.}
    \begin{tabular}{lccccccc}
    \hline\hline 
        region & 1 & 2 & 3 & 4 & 5 & 6 & 7 \\
        \hline
        Temperature (keV) & $4.00_{-0.36}^{+0.36}$ & $3.08_{-0.30}^{+0.31}$ & $3.19_{-0.32}^{+0.36}$ & $3.57_{-0.41}^{+0.45}$ & $3.19_{-0.33}^{+0.42}$ & $2.62_{-0.24}^{+0.33}$ & $3.12_{-0.28}^{+0.27}$  \\
        \hline
        Abundance & $0.99_{-0.30}^{+0.39}$ & $0.20_{-0.13}^{+0.16}$ & $0.31_{-0.16}^{+0.22}$ & $0.31_{-0.18}^{+0.23}$ & $0.30_{-0.16}^{+0.22}$ & $0.17_{-0.10}^{+0.14}$ & $0.68_{-0.21}^{+0.28}$ \\

    \hline
    \end{tabular}
    \label{tab:panda_label}
\end{table*}

\begin{table*}
    \centering
    \caption{The fitting results of the regions generated by \textit{Contour Binning}.}
    \begin{tabular}{lcccccccc}
    \hline\hline 
    region & 1 & 2 & 3 & 4 & 5 & 6 & 7 & 8 \\
    Temperature (keV) & $3.36_{-0.30}^{+0.37}$ & $3.18_{-0.30}^{+0.33}$ & $3.44_{-0.26}^{+0.27}$ & $3.04_{-0.27}^{+0.31}$ & $2.66_{-0.21}^{+0.26}$ & $3.09_{-0.36}^{+0.38}$ & $2.94_{-0.31}^{+0.36}$ & $2.81_{-0.37}^{+0.40}$ \\
    Abundance & $0.48_{-0.20}^{+0.28}$ & $0.25_{-0.13}^{+0.17}$ & $0.62_{-0.17}^{+0.20}$ & $0.42_{-0.17}^{+0.22}$ & $0.40_{-0.21}^{+0.32}$ & $0.39_{-0.16}^{+0.22}$ & $0.19_{-0.12}^{+0.21}$ & $0.30_{-0.11}^{+0.14}$ \\
    \hline
    region & 9 & 10 & 11 & 12 & 13 & 14 & 15 \\
    Temperature (keV)  & $2.01_{-0.22}^{+0.31}$ & $3.07_{-0.38}^{+0.46}$ & $2.99_{-0.28}^{+0.29}$ & $2.40_{-0.37}^{+0.45}$ & $2.36_{-0.34}^{+0.53}$ & $4.16_{-0.45}^{+0.52}$ & $3.53_{-0.40}^{+0.47}$ \\
    Abundance & $0.41_{-0.14}^{+0.18}$ & $0.10_{-0.07}^{+0.12}$ & $0.40_{-0.20}^{+0.29}$ & $0.21_{-0.14}^{+0.19}$ & $0.27_{-0.15}^{+0.20}$ & $1.55_{-0.56}^{+0.87}$ & $0.38_{-0.20}^{+0.28}$ \\
    \hline
    \end{tabular}
    \label{tab:contourbin_label}
\end{table*}

\end{document}